# Freeform Direct-write and Rewritable Photonic Integrated Circuits in Phase-Change Thin Films

Changming Wu[1], Haoqin Deng[1], Yi-Siou Huang[2,4], Heshan Yu[2,3], Ichiro Takeuchi[2], Carlos A. Ríos Ocampo[2,4] and Mo Li[1,5,*]

[1]Department of Electrical and Computer Engineering, University of Washington, Seattle, WA 98195, USA

[2]Department of Materials Science and Engineering, University of Maryland, College Park, MD 20742, USA

[3]School of Microelectronics, Tianjin University, Tianjin, 300072, China

[4]Institute for Research in Electronics and Applied Physics, University of Maryland, College Park, MD 20742, USA

[5]Department of Physics, University of Washington, Seattle, WA 98195, USA

[*]Corresponding author: moli96@uw.edu

## ABSTRACT

**Photonic integrated circuits (PICs) with rapid prototyping and reprogramming capabilities promise revolutionary impacts on a plethora of photonic technologies. Here, we report direct-write and rewritable photonic circuits on a low-loss phase change material (PCM) thin film. Complete end-to-end PICs are directly laser written in one step without additional fabrication processes, and any part of the circuit can be erased and rewritten, facilitating rapid design modification. We demonstrate the versatility of this technique for diverse applications, including an optical interconnect fabric for reconfigurable networking, a photonic crossbar array for optical computing, and a tunable optical filter for optical signal processing. By combining the programmability of the direct laser writing technique with PCM, our technique unlocks opportunities for programmable photonic networking, computing, and signal processing. Moreover, the rewritable photonic circuits enable rapid prototyping and testing in a convenient and cost-efficient manner, eliminate the need for nanofabrication facilities, and thus promote the proliferation of photonics research and education to a broader community.**

*Short Title:* Direct-write and rewritable photonic circuits

*One-sentence summary:* In a thin film of phase-change materials, photonic circuits can be directly written, erased, and modified by a laser writer.

**Introduction**

The application of photonic integrated circuits is rapidly spreading in diverse technology domains, ranging from computing(*1–3*), communication(*4–6*), sensing(*7–9*), and quantum technology(*10–13*). Traditionally, PICs are fabricated in thin film materials, including silicon(*4, 14, 15*), silicon nitride(*16, 17*), indium phosphide(*18, 19*), and lithium niobates(*20, 21*), using top-down nanofabrication processes including lithography, etching, and deposition on tools installed in cleanroom facilities. Compared with electronics, where prototypes can be rapidly built by students using discrete elements plugged into a breadboard, photonics research faces the barriers of the limited accessibility to fabrication facilities, the high costs associated with the fabrication process, and the extended design-to-device turnaround time. The environmental impact and chemical waste generated by the top-down fabrication process of PICs are also a concern. The combination of these barriers impedes widespread innovation and throttles the broader impact of photonics research and education.

Moreover, there is a growing interest in programmable PICs to realize highly flexible photonic networks for emerging applications, including optical computing, optical interconnect for neural network accelerators(*15, 22–24*), and quantum computing(*25–28*). Currently, programmable PICs are realized through a network of tunable components, including phase shifters, directional couplers, and interferometers, connected in a highly complex architecture(*29–31*). The programmability of those schemes has been limited to switching and rerouting the network, while freeform reconfiguration of the system's functionality remains difficult to achieve. As the complexity increases, programmable PICs face overwhelming challenges of scalability, programming precision, and flexibility, which limit their practicality and increase cost(*32–34*).

Here, we report a simple, fast turnaround and low-cost approach to creating and reprogramming PICs that could shift the paradigm in photonics research, prototyping, and education. By eliminating the reliance on traditional fabrication processes, our technique enables researchers to explore a wider space of design possibilities and system functionalities more rapidly. Moreover, such a technique will allow researchers and students who do not have access to nanofabrication facilities to prototype and reuse PIC designs and thus democratize photonics research to a broader community. It will enable more students and educators to engage in hands-on experimentation, thereby fostering innovation and knowledge dissemination and generating a broader impact to promote workforce development in photonics.

Our technique is based on direct laser writing (DLW) on phase-change material (PCM) thin films. The method writes the PICs in only one optical patterning step and without using any traditional lithography and etching processes. The PICs are created by utilizing PCM's dramatic refractive index contrast between the two nonvolatile phases, amorphous and crystalline, which are reversibly switchable using optical pulses[35]. We have demonstrated direct writing of functional PICs consisting of a full package of building-block photonic components, including waveguides, gratings, ring resonators, couplers, crossings, and interferometers. Remarkably, these direct-write PICs are also rewritable, allowing convenient erasure and recreation of the circuits either partially or entirely, thereby completely changing their functionality to suit very different application scenarios.

Previously, laser writing to control phase transition in PCMs has been used in optical storage, such as rewritable compact disks (CD-RW)[35]. Free-space optical switching of PCMs has also been used to realize reconfigurable metasurface[36, 37], programmable silicon photonics[38], rewritable color displays[39], and polariton nanophotonic devices[40]. Laser patterning of photonic circuits has been reported, including femtosecond laser writing of waveguides in silica [41–45] and chalcogenide glasses [47–49] and a programmable III–V semiconductor-based photonic device[46]. However, the photon-induced refractive index contrast for glass waveguide application is small, typically less than 0.1[47–49]. This limitation constrains not only the footprint of the devices but also their potential applications. On the other hand, optical patterning in III-V semiconductors is volatile, necessitating continuous laser irradiation for sustainability. Additionally, none of the above techniques creates complete end-to-end PIC systems with multiple reconfigurable functions, as we report here.

**Results**

**Direct-write and rewritable PICs**

Fig. 1A illustrates the concept of direct writing, erasing, and rewriting PICs. The PIC is written on a standard oxidized silicon substrate, which is coated with a 200-nm-thick $SiO_2$ layer covering a 30-nm-thick $Sb_2Se_3$ layer on a 330-nm-thick $Si_3N_4$ film. The $SiO_2$ capping layer protects and prevents oxidation of the $Sb_2Se_3$ layer. Waveguiding in the $Sb_2Se_3$ film is achieved by using the crystalline phase ($cSb_2Se_3$) as the high-index core and the amorphous phase ($aSb_2Se_3$) as the cladding (Fig. 1B). This binary phase configuration, that is, no mixed phases, enables the confinement of the fundamental transverse electric ($TE_0$) optical mode within a $cSb_2Se_3$ waveguide, assisted by the underlayer of $Si_3N_4$, as shown in the simulated mode profile in Fig. 1C. We directly write the circuit layout on a blank $cSb_2Se_3$ thin film using a commercial laser writing system (Heidelberg DWL 66+, 405 nm laser), mainly to leverage its precision stage and computer control system for large area writing. Alternatively, a homebuilt system using an off-the-shelf laser diode is sufficient to write a smaller area (Supplementary Note 2). To write the PICs, the focused laser beam with a power of 27.5 mW is scanned across the film at a speed of 3.0 $mm^2$/min, inducing a controlled phase transition from $cSb_2Se_3$ to $aSb_2Se_3$ (Supplementary Note 2) to create the claddings and define the core. This laser-controlled phase transition creates a circuit with a resolution limited by optical diffraction. Fig. 1D shows a series of rectangular $aSb_2Se_3$ structures created by DLW with varying widths ranging from 1 µm to 200 nm. The minimum achievable feature size is 300 nm, exceeding the resolution reported in previous works[37–40]. Moreover, the circuit layout can be erased either locally or globally. Local erasure is achieved by scanning the laser beam at a lower speed (0.1 $mm^2$/min) with a reduced laser power of 15 mW, inducing a phase transition from $aSb_2Se_3$ back to $cSb_2Se_3$ in desired areas (Fig. 1E). This capability enables modifications and corrections to the PIC design. Alternatively, global erasure can be accomplished by heating the whole substrate to above 180 °C, promoting a complete phase transition across the entire PCM film. Thereby, the DLW technique provides very versatile capabilities for writing and rewriting phase-change PICs in freeform without using any advanced fabrication tools.

**Direct-write PIC Components**

We first demonstrate the DLW of high-quality building-block components of PICs. The first example (Fig. 2A) is a racetrack ring resonator coupled with a bus waveguide connected to input and output grating couplers, all directly written with DLW. The resonator has a width of 1.2 µm and a radius of 120 µm with a gap of 350 nm to the bus waveguide (Fig. 2B). The pair of grating couplers couple light from optical fibers into the bus waveguide (Fig. 2C) with an efficiency of -14 dB. Fig. 2D shows the measured transmission spectrum through the bus waveguide, showing the resonances of the ring resonator with an intrinsic quality factor is $Q_i \sim 12{,}700$, corresponding to a propagation loss of 2.8 dB $mm^{-1}$. The loss in the $cSb_2Se_3$ waveguide is mainly attributed to the scattering induced by the grain boundaries in the waveguide, as in-plane crystalline grains within the cSb$_2$Se$_3$ region can be observed with high-resolution electron microscopy (Supplementary Note 3). These grains, typically a few microns in size, are also randomly oriented(*50*). Because of the optical birefringence of cSb$_2$Se$_3$(*39*), these randomly oriented grains result in the uneven absorption of laser energy during the DLW and limit the smoothness of the waveguide. The scattering loss can be mitigated when a lower erasure temperature is used to form larger crystal grains with sizes on the order of tens of microns, which we have observed. Further mitigation of the loss will require improved optical phase-change materials with finer polycrystalline structures. Nevertheless, the propagation loss of the cSb$_2$Se$_3$ waveguide is consistent with the reported values of dielectric waveguides integrated with Sb$_2$Se$_3$(*50–53*) .

The devices written on phase-change thin film exhibit consistent and reliable performance. To demonstrate this, we write multiple Mach-Zehnder interferometers (MZIs) with varied path length differences, $\mathit{\Delta L}$, between the two arms (Fig. 2E). The measured transmission spectrum is shown in Fig. 2F. The consistently high extinction ratio (> 20 dB) indicates the 50/50 beam splitters (inset, Fig. 2E) performs very well. Fitting the spectra, we extract a group index $n_g = 2.47$ for a 1.2-µm-wide Sb$_2$Se$_3$ waveguide, which agrees well with the simulated value of 2.43. The characterization of other photonic building block elements, including directional couplers, waveguide crossings, and inverse-designed bends with low loss, are included in Supplementary Note 4.

**Programmable interconnect fabric and crossbar matrix**

Programmable optical interconnect fabric is a crucial photonic architecture to enable reconfigurable connectivity in telecommunication, computing, and network systems(*54, 55*). Using the DLW technique, we can create and, on-demand, route and reroute such interconnect fabric. In Fig. 3A–C, we demonstrate a 3×3 array, which establishes connectivity between three input and output ports using cSb$_2$Se$_3$ waveguides. We define the connection configuration of the fabric using a transmission matrix **M** in $S_{out}$=**M · **$S_{in}$, where $S_{out}$ and $S_{in}$ represent the output and input intensity vectors, respectively. As shown in Fig. 3A, we initially patterned the fabric to connect input/output ports in a configuration of 1→2, 2→1, and 3→3. The measurement result of this transmission matrix **M** shows that the extinction ratio between desired and undesired connections exceeds 20 dB (Fig. 3D). We then reroute the fabric by erasing all the waveguides in the designated connection region (Fig. 3B) and subsequently rewriting the waveguides to establish new connections with an updated configuration. As illustrated in Fig.3C, we reroute the switch array to the configuration of 1→2, 2→3, and 3→1. The measured **M** has changed accordingly but maintains a high extinction ratio >18 dB. More data on routing and rerouting the optical fabric with other connection configurations can be found in Supplementary Note 6. Additionally, we observe a residual image from the previous writing even after completely erasing the pattern. The residual image does not impact the performance of the newly patterned devices, though. We argue that this residual imprint can be explained by the optical birefringence resulting from size and orientation variations in $cSb_2Se_3$ crystal grains during the annealing processes. Specifically, during the laser erasing process, the crystal grains grow in alignment with the moving direction of the laser spot, while no preferred growth direction is present during the thermal erasing process[50].

A more complicated transmission matrix can be achieved by using a photonic crossbar array, which is a universal architecture for implementing multiply-accumulate (MAC) operation in linear optical computing[56, 57]. Fig. 3F displays the optical image of a 14×14 photonic crossbar array written on the PCM thin film. As proof of concept, we demonstrate the operation of a 3×3 sub-array to represent a programmable MAC core. The input vector is encoded by the intensities of optical signals in multiple wavelength channels, which are sent into each row of the crossbar array, respectively. As illustrated in Fig. 3G, each unit cell of the crossbar array consists of a row waveguide, a column waveguide, and a cross-coupling waveguide, which extracts the input signal from the row waveguide and couples it to the column waveguide through optimized directional couplers. Supplementary Note 4 provides data on the splitting ratio of the directional coupler with various coupling lengths. The output of each column waveguide is the weighted sum of the input signal at each row and is measured incoherently by a photodetector, thus performing the MAC operation for matrix-vector multiplication (MVM). Each matrix element is programmed in the transmission of each unit cell's cross-coupling waveguide (see Fig. 3F). The output vector is obtained by measuring the total output power in each column of the output channels. A more detailed elucidation of crossbar array implementation can be found in previous works[56, 57].

For demonstration, we initially designed and direct-wrote the 3×3 crossbar array to equally distribute the input power into the output to represent the matrix $\begin{bmatrix} 1 & 1 & 1 \\ 1 & 1 & 1 \\ 1 & 1 & 1 \end{bmatrix}$. However, due to writing imperfections (similar to fabrication variation in conventional PICs), the matrix represented by the crossbar array exhibits errors, as shown in the measured results in Fig. 3I. Remarkably, the DLW technique makes it straightforward to adjust each element by locally modifying the cross-coupling waveguides, thereby correcting the error to restore the designed functionality (Fig. 3J). It is also easy to reprogram the matrix in a grayscale-like fashion by erasing or restoring a section of the cross-coupling waveguide. As an example, we can set the element at the center and four corners to 0 (Fig. 3K) and decrease the center matrix element to 0.5 (Fig. 3L) to represent the matrix $\begin{bmatrix} 0 & 1 & 0 \\ 1 & 0.5 & 1 \\ 0 & 1 & 0 \end{bmatrix}$. Further data on tuning the attenuation and phase response of the $Sb_2Se_3$ waveguide can be found in Supplementary Note 5.

**Shaping Spectral Response of Coherent PICs**

Coherent PICs universally use interferometric and resonant components and require precise spectral responses for filtering and coherent signal processing(*58–60*). Because of inevitable fabrication imprecision, it is necessary to shape these components' spectral responses to meet these requirements. The DLW technique allows us to incrementally trim the spectral response of a coherent optical filter step-by-step, offering a highly controlled method for fine-tuning its performance. In addition, we can even add new coherent components to change the spectral response entirely. As shown in Fig. 4A, we first write an MZI, which has a typical transmission spectral response as measured in Fig. 4G. We then erase a section of the lower arm of the MZI (Fig. 4B) and subsequently restore it and add a racetrack ring resonator, as depicted in Fig. 4C. The resulting transmission spectrum of the ring coupled-MZI displays features the characteristics of both the MZI and the ring resonator. Afterward, we erase portions of both the ring resonator and the MZI (see Fig. 4D) and rewrite the circuit to a double-injection ring (DIR) filter(*61*). In the DIR, the output light consists of cumulative contributions from two nominally identical add-drop ring resonators. The first contribution comes from the drop-port of the ring, with the input light incident from the lower arm of the original MZI. The second contribution originates from the through-port of the ring, with the light incident from the upper arm of the original MZI. As a result, the DIR exhibits a substantial different transmission spectrum with a much larger FSR (Fig. 4G). Finally, we can trim the DIR spectral response by modifying its parameters. We adjust the coupling between the output waveguide and the racetrack ring resonators. This is done by widening the gap between the waveguides from 350 nm to 500 nm and reducing the waveguide width (Fig. 4F). The transmission spectrum of the modified DIR (DIR2 in Fig. 4F) changes as expected, thus demonstrating the spectral tunability of the filter. We note that all measured spectra from the filter agree well with the numerical models (further details are provided in Supplementary Note 7). This provides additional evidence of the stability and reliability of writing and rewriting the phase-change photonic circuits.

**Discussion**

In conclusion, we have presented a flexible, reliable, and cost-effective technique for directly writing and rewriting photonic circuits on low-loss phase-change thin films. This technique simplifies the complex nanofabrication processes typically required for fabricating integrated photonic devices down to one-step laser writing and enables reusing the same chip/die. Although we have utilized a commercial laser writer in this work, we emphasize that the same results can be achieved with a much lower-cost laser writing system, which incorporates a laser diode, a focusing objective, a high-precision motion stage, and a computer control system (refer to Supplementary Note 2 for further information). This approach leverages the versatility of the DLW and the nonvolatile, low-loss, and high-contrast properties of the phase-change material, offering an unprecedented level of flexibility. We have demonstrated PICs consisting of a full package of elementary photonic components, including waveguides, grating couplers, ring resonators, MZIs, programmable optical switch fabrics, reconfigurable photonic crossbar array, and tunable optical filters. Although our demonstrations have been conducted in a near-*in-situ* fashion—in steps of writing, measuring, modifying, and checking—real-time reconfiguration and feedback-controlled adaptation(*6, 30, 46, 62*) of the PICs are entirely feasible.

Furthermore, the application scenario can be expanded by introducing a multi-level grayscale design(*37, 39, 57, 63–65*) instead of the current binary design or by selecting appropriate substrates tailored to specific applications. For example, the phase-change thin films can also be integrated on LiNbO$_3$-on-insulator substrates, enabling the development of programmable electro-optical or acoustic-optical circuits. These advantages underscore the potential of the DLW technique in enabling the rapid prototyping and testing of innovative photonic circuits, using only a low-cost tool that is affordable to a wide range of research and education communities.

**Materials and Methods**

**Substrate preparation**

To prepare the substrate for direct laser writing, a 30nm-thick $Sb_2Se_3$ thin film sputtered on a 330-nm-thick $Si_3N_4$ on an oxidized silicon substrate at room-temperature. The whole substrate was then capped with a 200-nm-thick $SiO_2$ for protection. After these two steps, the substrate was ready for laser writing.

**Laser writing setup**

We utilize a commercial 405 nm direct writing lithography tool (Heidelberg DWL66+ with its Hires laser head) for the writing of the phase-change PICs. The laser power is set to 27.5 mW with a scanning speed of 3 $mm^2$/min. Additionally, a homebuilt laser writing system is also applied to local tune, erase, and rewrite the PICs (see Fig. S3). In the homebuilt system, a 637 nm laser diode with a maximum output power of 170 mW (HL63133DG) is focused onto the substrate using a 50× objective lens with an NA of 0.55. To write a single spot, a rectangular pulse of 200 ns pulse duration and pulse energy of 50 nJ is used. To erase a single spot, the pulse duration is set to 5 ms with a pulse energy of 333 µJ. The sample was mounted on a 2D x-y closed-loop motion stage, capable of a minimum moving step size of 50 nm. By synchronizing the controlled light pulse with the movement of the substrate on the stage, the desired pattern can be achieved. More details on the optical setup can be found in Supplementary Note 2.

**Figures**

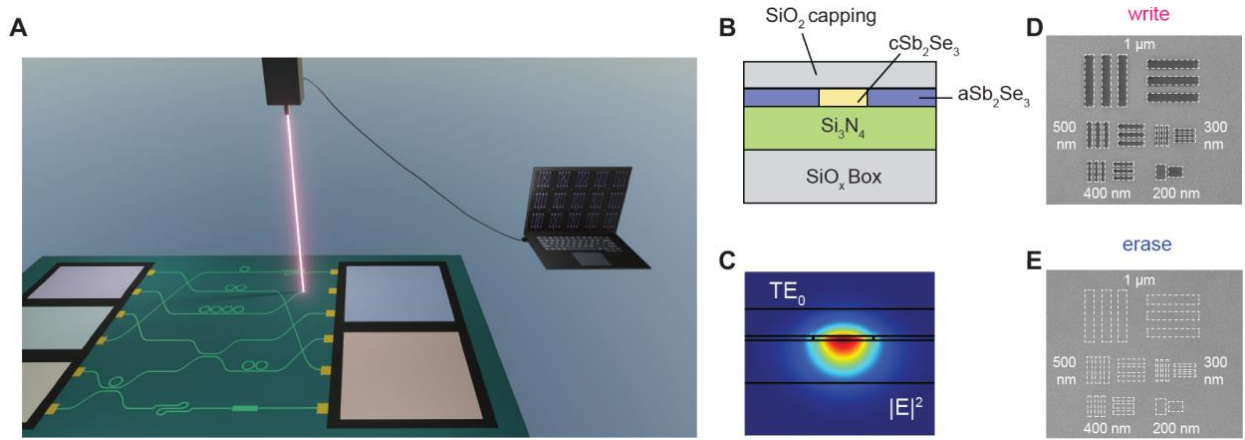

**Fig.1. Direct-write and rewritable phase-change photonic integrated circuits.** (**A**) Artistic illustration of freeform writing and rewriting PICs on Sb₂Se₃ thin film. (**B**) The cross-sectional view of a cSb₂Se₃ optical waveguide structure. The waveguide is directly written in the PCM thin film without fabrication processes. (**C**) The simulated $/E/^2$ profile of the TE₀ mode in a cSb₂Se₃ waveguide, which is 1.2-μm-wide, 30-nm-thick, sits on top of a 330-nm-thick Si₃N₄-on-insulator substrate, and is capped with a 200-nm-thick SiO₂ for protection. (**D**) Optical image of aSb₂Se₃ resolution test patterns written on cSb₂Se₃ thin film. The minimum feature size achieved is 300 nm. (**E**) The same test pattern as in (D) is erased back to cSb₂Se₃.

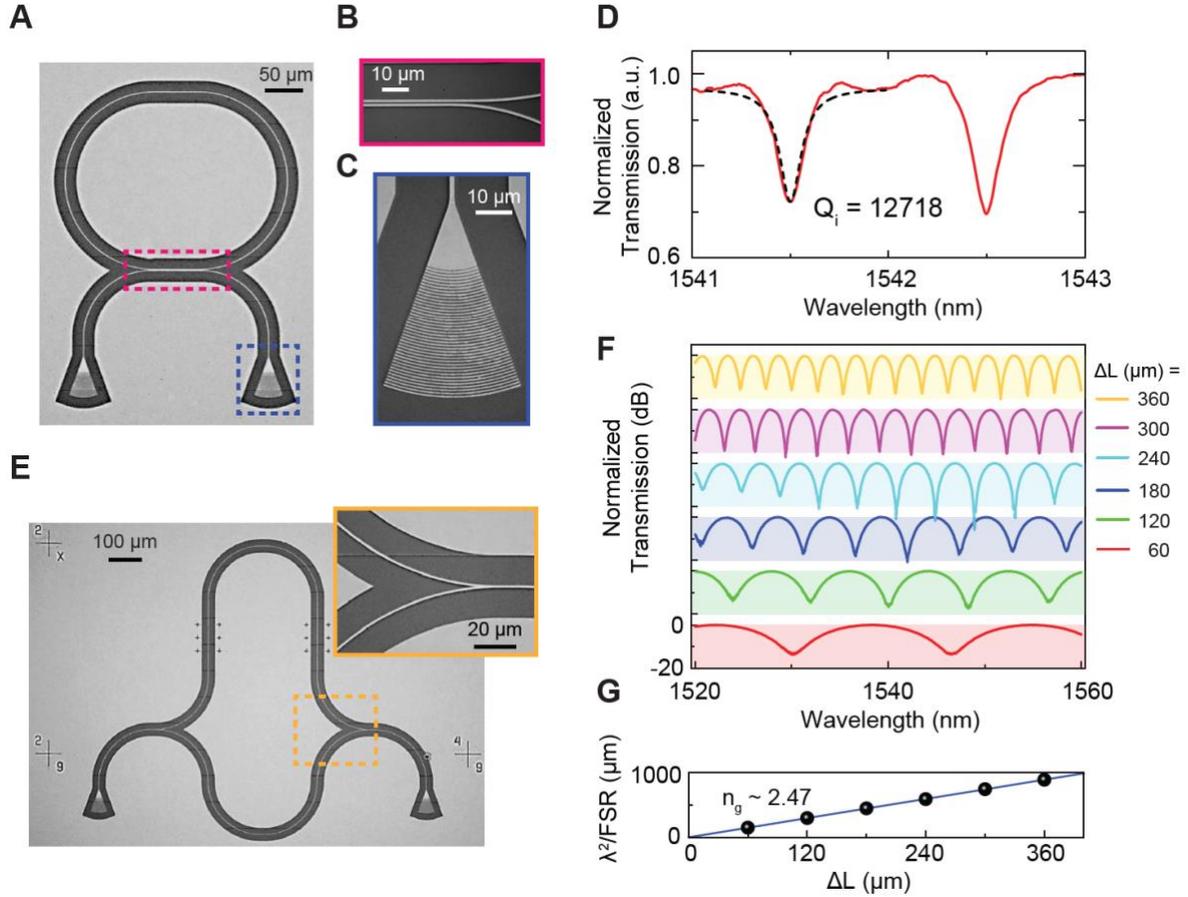

**Fig.2. Direct-write photonic components and their characterization.** (**A**) Optical image of a racetrack ring resonator that is directly written on the Sb₂Se₃ thin film. (**B**) and (**C**) The zoomed-in optical image of the waveguide-ring coupling region (B) and the grating coupler (C). (**D**) The transmission spectrum of the ring resonator as shown in (A). The spectrum is normalized to the spectrum of a pristine Sb₂Se₃ waveguide. The intrinsic *Q* factor is 12,718. (**E**) Optical image of a direct-write PCM Mach-Zehnder interferometer (MZI). Inset: the zoomed-in image of the Y-combiner part in the MZI. (**F**) Normalized transmission spectra of the MZIs. Spectra have been vertically offset for clarity. (**G**) $\lambda^2$/free-spectral-range vs. the length difference between two arms of the MZI, $\varDelta L$. Linear fitting yields the waveguide group index $n_g$ to be 2.47.

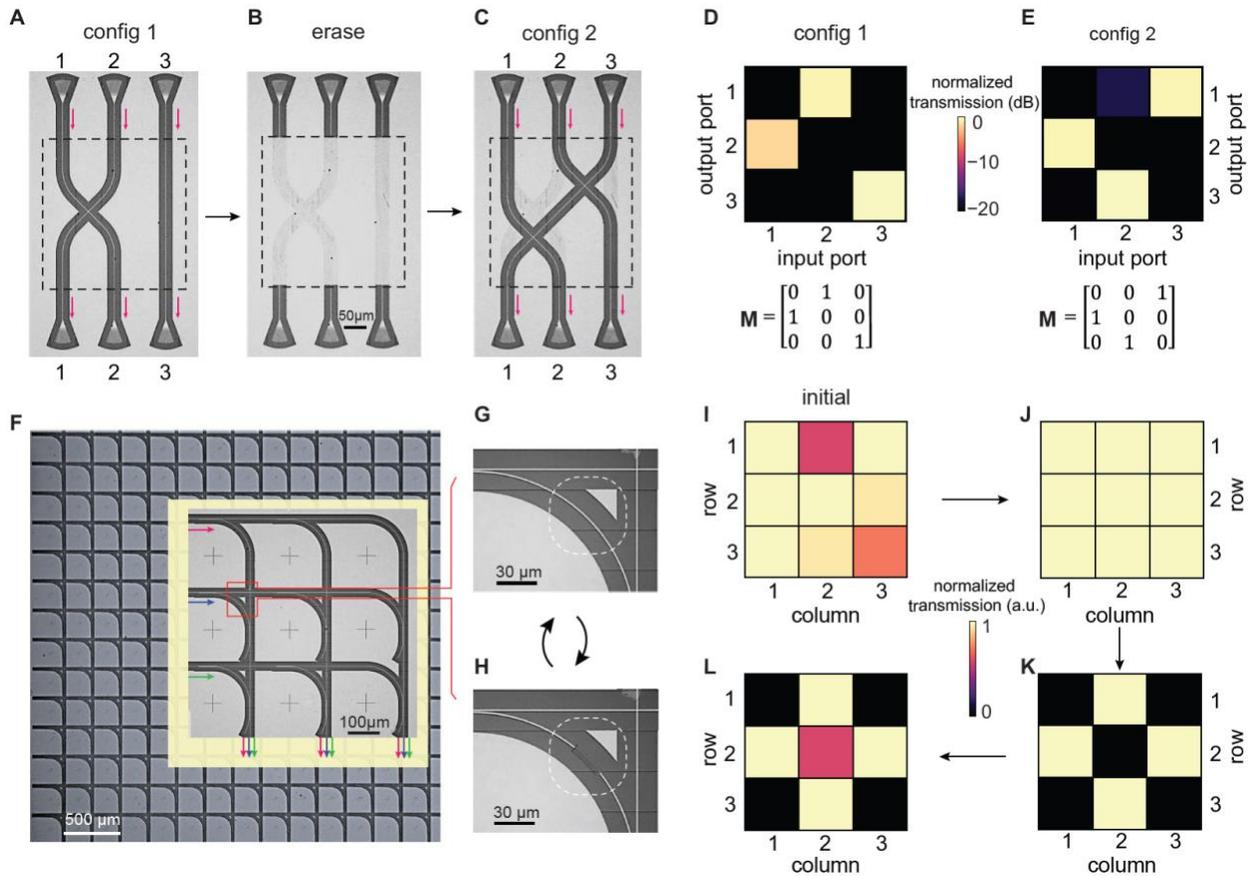

s

**Fig.3. Programmable photonic switch array and crossbar array.** (**A**) Optical image of a 3×3 optical switch array with the initial connection configuration (matrix 1). The connection region of the optical switch is then erased (B) and rewritten (C) into a new connection configuration (matrix 2). (**D**) and (**E**) The measured transmission matrix of switch 1(D) and switch 2 (E), respectively. (**F**) Optical image of a DLW 14×14 crossbar array. Inset: the 3×3 sub-array used for testing. To decrease or increase the transmission of a specific crossbar element, the cross-coupling waveguide is partially erased (H) or recovered (G), respectively. (**I-K**) The transmission matrix of the crossbar array after each configuration step. The crossbar array is initially designed to equally distribute the input power into outputs. Due to writing imperfections, the transmission matrix of the crossbar array has errors (**I**) and is corrected by DLW to restore the designed functionality (J). (**K** and **L**) Transmission matrix after setting the element at the center and four corners to 0 (K) and then resetting the center matrix element to 0.5 (L).

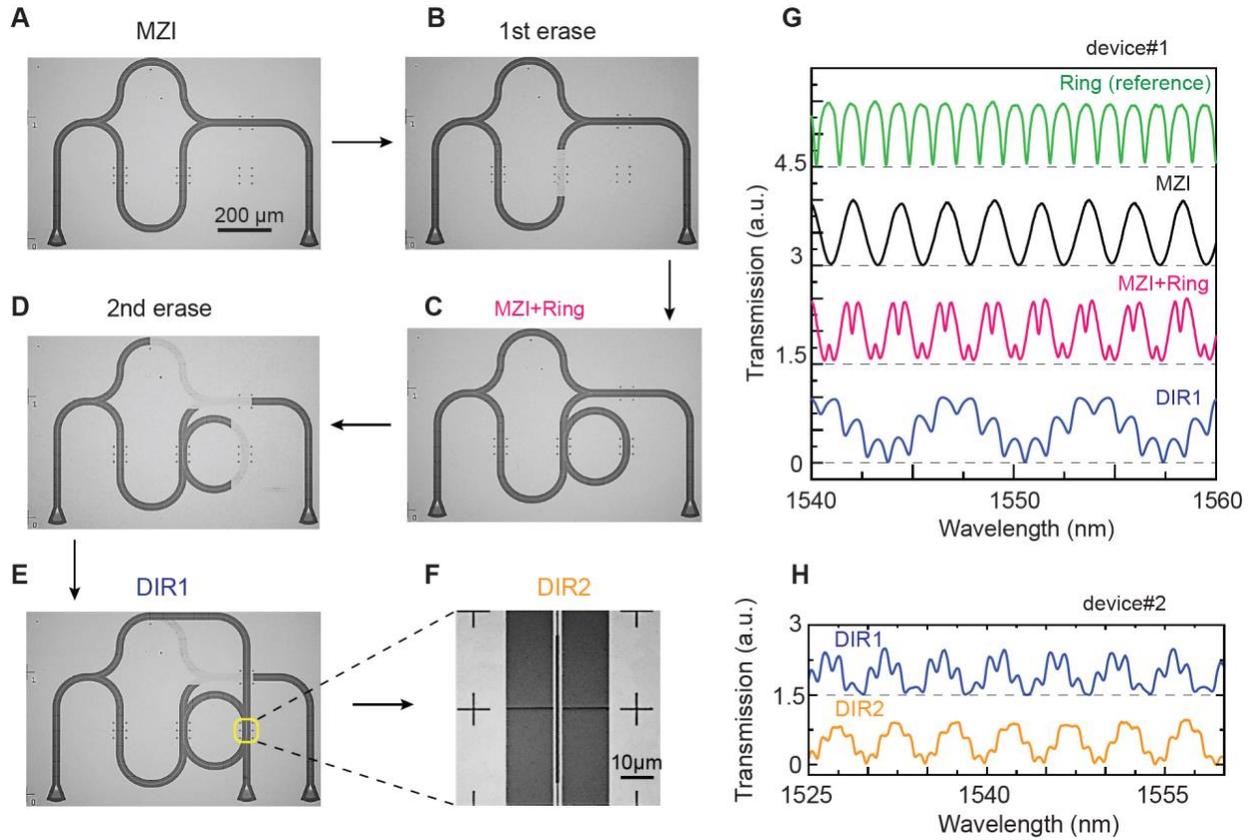

**Fig.4. Shape the spectral response of an optical filter in steps.** (**A**) An MZI written by DLW on a Sb$_2$Se$_3$ thin film. (**B**) Erase part of the bottom arm of the MZI. (**C**) Add a ring resonator to the erased region of the MZI. (**D**) Erase part of the ring resonator and the MZI. (**E**) Reconnect the device as the double-injection ring (DIR) filter. (**F**) Tuning the coupling by increasing the gap between the ring resonator and the top arm of the Y-splitter. (**G**)The transmission spectra of the optical filter after each reconfiguration step, respectively. Spectra have been vertically translated for clarity. (**H**)The transmission spectra of the double-injection ring filter before (blue curve) and after (orange curve) tuning the ring resonator coupling.

**Acknowledgment**

**Funding**: C.W., H.D., and M.L. acknowledge the funding support provided by ONR MURI (award no. N00014-17-1-2661) and the National Science Foundation (award no. CCF-2105972). Y.H., H.Y., I.T., and C.O. acknowledge the funding support by NSF (award no. ECCS-2210168 and DMR-2329087). Part of this work was conducted at the Washington Nanofabrication Facility / Molecular Analysis Facility, a National Nanotechnology Coordinated Infrastructure (NNCI) site at the University of Washington with partial support from the National Science Foundation via awards NNCI-1542101 and NNCI-2025489.

**Author contributions:** M.L. and C.W. conceived the research and designed the experiments. C.W. and H. Q. designed and fabricated the photonic devices and performed the experiments. H.Y., Y.H., C.R. and I.T. developed and deposited the $Sb_2Se_3$ thin film. C.W., H.Q., and M.L. analyzed the data. C.W. and M.L. wrote the manuscript with contributions from all the authors. All authors discussed the results and commented on the manuscript.

**Competing interests:** M.L., C.W., and H.Q. have filed a U.S. Provisional Patent Application (No.: 63/516,267) titled "Freeform Direct-Write and Rewritable Photonic Circuits on Phase-Change Thin Layers" to the U.S. Patent and Trademark Office (USPTO) on July 28, 2023. The authors declare no other competing interests.

**Data and materials availability:** All data needed to evaluate the conclusions in the paper are present in the paper and/or the Supplementary Materials.

# Freeform Direct-write and Rewritable Photonic Integrated Circuits in Phase-Change Thin Films

Changming Wu[1], Haoqin Deng[1], Yi-Siou Huang[2,4], Heshan Yu[2,3], Ichiro Takeuchi[2], Carlos A. Ríos Ocampo[2,4] and Mo Li[1,5]

[1]*Department of Electrical and Computer Engineering, University of Washington, Seattle, WA 98195, USA*

[2]*Department of Materials Science and Engineering, University of Maryland, College Park, MD 20742, USA*

[3]*School of Microelectronics, Tianjin University, Tianjin, 300072, China*

[4]*Institute for Research in Electronics and Applied Physics, University of Maryland, College Park, MD 20742, USA*

[5]*Department of Physics, University of Washington, Seattle, WA 98195, USA*

Corresponding author: moli96@uw.edu

# Supplementary Note 1. Substrate preparation and phase-change PICs parameters

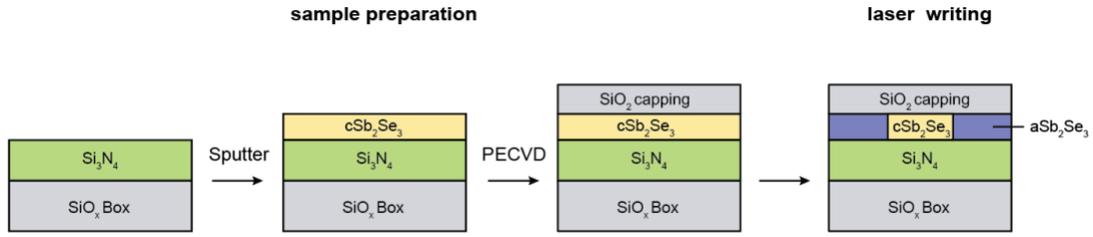

**Figure S1**. **Process flows for direct laser writing of PICs.** Only two steps, Sb₂Se₃ sputtering and SiO₂ deposition, are required for sample preparation. PECVD: plasma-enhanced chemical vapor deposition.

### Table T1. Parameters of the phase-change PICs

| | | | |
|---|---|---|---|
| SiO₂ capping thickness (nm) | 200 | Sb₂Se₃ thickness (nm) | 30 |
| Si₃N₄ thickness (nm) | 330 | SiO₂ box thickness (μm) | 2.8 |
| Sb₂Se₃ grating coupler period (nm) | 890 | Grating coupler duty cycle | 0.6 |
| Sb₂Se₃ waveguide width (μm) | 1.2 | Waveguide Bending radius (μm) | 120 |
| Directional coupler gap distance (nm) | 350 | Ring resonators radius (μm) | 120 |

**Supplementary Note 2. Direct laser writing setup**

We employ a commercial direct writing lithography tool (Heidelberg DWL66+ with its Hires laser head) for the writing of the phase-change PICs (Fig. S2). We emphasize the advantages of utilizing this commercial laser writing tool as it provides both high writing speed and resolution, enabling efficient and precise fabrication of photonic structures. The tool utilizes a continuous wave (CW) 405 nm diode laser as the laser source. An acousto-optic modulator (AOM) is employed to regulate and modulate the laser intensity. Subsequently, the laser beam passes through an acousto-optic deflector (AOD) and is focused onto the sample surface using a high numerical aperture (NA) lens. Due to AOD's limited deflection angle, this tool's maximum writing field is 60 µm. To fully extend the patterning range, the sample is placed on a high-speed and high-precision 2D motion stage. By incorporating both the AOD and the motion stage, the system enables the fast writing of PICs design on a phase-change thin film with a minimum feature size of 300 nm. In our experiment, we set the laser power to 27.5 mW, respectively, with a laser scanning speed of 3 mm²/min. Furthermore, another key benefit of the tool is its compatibility with commonly used patterning software. Specifically, the tool can directly load designs in formats such as GDS or CAD and transfer the design into patterns, eliminating the need for additional pattern conversion procedures.

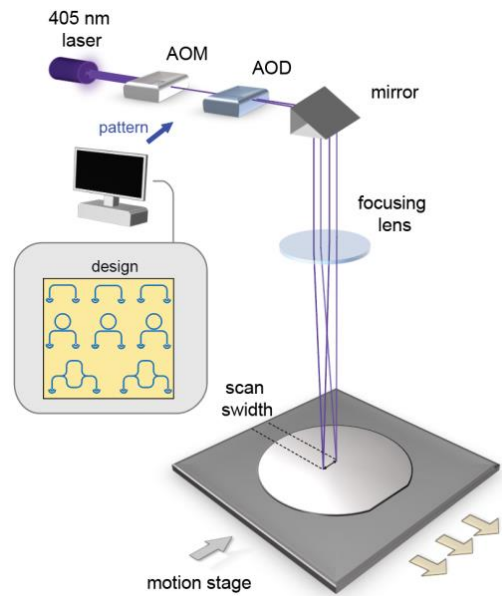

**Figure S2. Schematic diagram of the direct laser writing experimental setup.** AOM: acousto-optic modulator. AOD: acousto-optic deflector.

While using a commercially available direct laser writer is convenient, we note that it is possible to construct a homebuilt laser writing system to minimize equipment requirements. To this end, we also demonstrate the approach that utilizes a cheap, high-power laser diode as the source for laser writing.

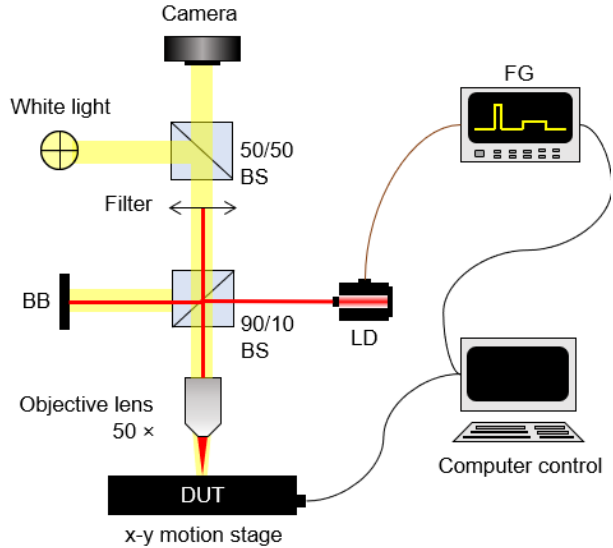

**Figure S3**. **Schematic diagram of the homemade direct laser writing setup.** LD: laser diode. BS: beam splitter. BB: beam block. FG: function generator. DUT: device under test.

As illustrated in Fig. S3, a 637 nm laser diode with a maximum output power of 170 mW (HL63133DG) is utilized for the writing and erasing processes. Precise control of laser pulse parameters is achieved by modulating the diode current with a function generator. The modulated laser output is then focused onto the substrate using a 50× objective lens with an NA of 0.55. To write a single spot, a rectangular pulse of 200 ns pulse duration and pulse energy of 50 nJ is used. To erase a single spot, the pulse duration is 5 ms with a pulse energy of 333 µJ. Both the writing and erasing processes achieved a diffraction-limited resolution of 500 nm, which can be further improved using a laser diode with a shorter wavelength. To facilitate precise positioning of the substrate, a 2D x-y closed-loop motion stage is used, which is capable of achieving a minimum moving step size of 50 nm. By combining the controlled light pulse with the movement of the substrate on the stage, the desired pattern was achieved. Moreover, for faster erasure of the pattern over a larger area, the laser diode was operated in continuous wave (CW) mode at a fixed power of 15 mW while scanning the stage at a speed of 0.1 mm$^2$/min. This method effectively erased all patterns in the path of the laser spot.

We note that in this work, the writing of PICs is performed using the commercial Heidelberg DLW66+ writer, while most of the local tuning and erasing of the PICs is achieved by the homebuilt system.

**Supplementary Note 3.  Crystal grains in crystalline Sb$_2$Se$_3$ thin film**

Fig. S4a shows the differential interference contrast (DIC) optical microscopic image of a Sb$_2$Se$_3$ photonic waveguide. The randomly oriented crystalline grains with sizes ranging from 2 to 5 µm are observed in both the waveguide and grating couplers regions. These grains likely contribute to additional scattering losses, which can limit the performance of phase-change PICs. To gain a better understanding and control over the formation of these crystalline grains, DIC microscopic images of Sb$_2$Se$_3$ thin films under various annealing conditions are presented in Fig. S4b to S4d. To ensure the Sb$_2$Se$_3$ thin films were in their crystalline phase, they were initially thermally annealed at 180℃ for 10 minutes. Following annealing, similar polycrystalline grains were observed. Subsequently, a 100 µm × 100 µm aSb$_2$Se$_3$ square was written on the crystalline thin film, as shown in Fig. S4b. The sample was then baked at 160℃ for 5 minutes, resulting in Fig. S4c. Interestingly, as the annealing temperature was reduced, the cSb$_2$Se$_3$ grains exhibited a preference for growth along a direction perpendicular to the boundary between aSb$_2$Se$_3$ and cSb$_2$Se$_3$. These grains exhibited a significantly larger domain size, ranging from 10 to 20 µm. In addition to the thermally annealed samples, laser-annealed samples were also obtained, as shown in Fig. S4d. Unlike the thermal annealing process, the laser annealing resulted in a distinct crystal structure, where the crystal grain growth was guided by the movement of the laser spot. Consequently, the large crystalline regions exhibited directional growth that was associated with laser movement. We note that the difference in size and orientation of cSb$_2$Se$_3$ crystal grains resulting from thermal (at 180℃) and laser annealing processes contribute to the observed residual effects in Fig. 3 and Fig. 4 after rewriting of phase-change PICs.

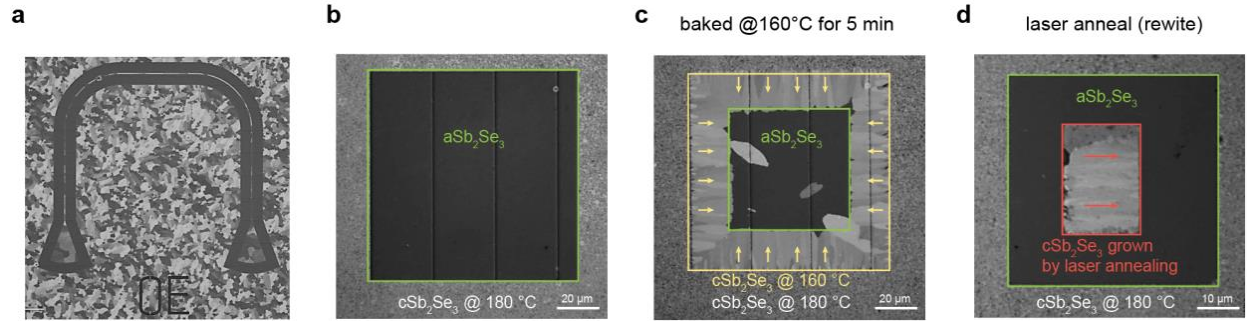

**Figure S4**. **DIC microscopic images of Sb₂Se₃ thin film device and samples. a**. DIC image of a phase-change photonic waveguide, the randomly oriented crystalline grains are observed in the waveguide and grating couplers regions, leading to additional scattering loss. **b and c**. An aSb₂Se₃ square is written on cSb₂Se₃ (annealed at 180°C) (a) and then baked at 160°C for 5 minutes (b). The yellow arrows indicate the grain growth direction is perpendicular to the aSb₂Se₃/cSb₂Se₃ boundary. **d**. cSb₂Se₃ grown on aSb₂Se₃ region through laser annealing (rewrite) process. The cSb₂Se₃ grains are grown along laser-moving directions as indicated by the red arrows. Green box: aSb₂Se₃ region. Yellow box: cSb₂Se₃ grown at 160°C. Red: cSb₂Se₃ grown by laser annealing.

The propagation loss in the waveguide is strongly mitigated when the crystal grains in the waveguide are larger. To show this, we compare the transmission spectra of waveguides with small and large gains.

The waveguide with small grains is directly written on the cSb₂Se₃ thin film with small grains. This substrate is initially prepared by baking the aSb₂Se₃ substrate on a hot plate at 180 °C. This process induces crystallization, forming cSb₂Se₃ with small grains.

On the other hand, creating the waveguide with larger grains undergoes a three-step process. We started with a cSb₂Se₃ thin film with small grains. Firstly, both the waveguide and the cladding areas are converted to the amorphous phase by laser writing. Secondly, the written pattern is erased at 160 °C for 10 minutes. This thermal annealing process leads to the growth of larger grains along a direction perpendicular to the boundary between aSb₂Se₃ and cSb₂Se₃ in the previously patterned area. Figure S5, captured with a differential interference contrast (DIC) mode microscope, show the effect. As shown in Fig. S5a, when observed using a filter with polarization directed along the vertical waveguide (polarization 1), the vertical part of the waveguide appears dark (highlighted within the red circle), while the horizontal part of the waveguide appears bright (highlighted with the green circle). Upon rotating the filter's polarization direction by 90 degrees (polarization 2), the bright and dark contrast reverses (see Fig. S5b). The polarization contrast stems from the birefringence of the crystalline phase. The uniform contrast, within the horizontal and vertical sections, indicates large crystalline grains. Finally, we write the waveguide within this prepared cladding areas so that the waveguide core consists of cSb₂Se₃ with much larger grain than the previous method.

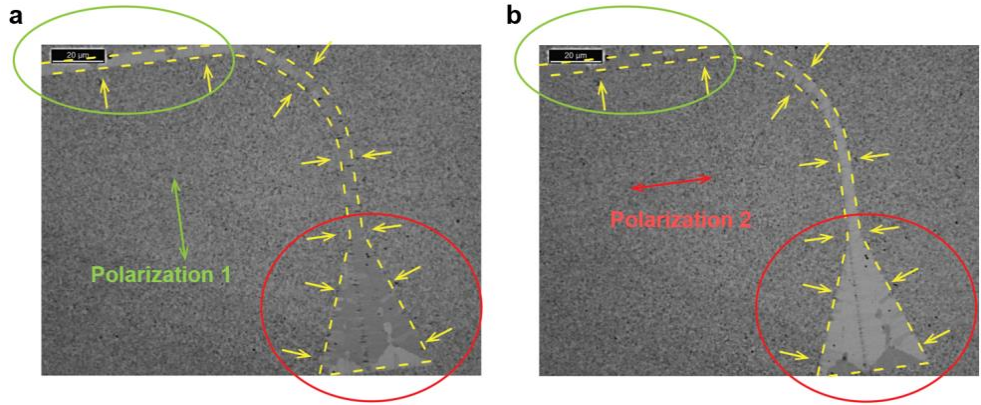

**Fig. S5 The DIC optical images of a waveguide cladding region after baking at 160 °C. a**. The DIC optical image captured using a filter with polarization aligned with the vertical waveguide. **b**. The DIC image of the same device captured after the filter's polarization rotating 90 degrees. This thermal annealing process leads to the growth of larger grains along a direction perpendicular to the boundary between aSb₂Se₃ and cSb₂Se₃ in the previously patterned area, as indicated by the yellow arrow.

The large grains in the waveguide core lead to reduced scattering between the crystalline grains and the birefringence effect. This is due to the growth of grains perpendicular to the waveguide direction (see yellow arrows in Fig. S5), resulting in an aligned crystal orientation in relation to the light traveling within the waveguide. We measured the transmission spectra of identical waveguides with small and large grains. Our measurements shown in the Fig.S6 demonstrate a significant reduction in propagation loss, by ~12 dB, in the waveguide with larger grains as compared to the waveguide with smaller grains.

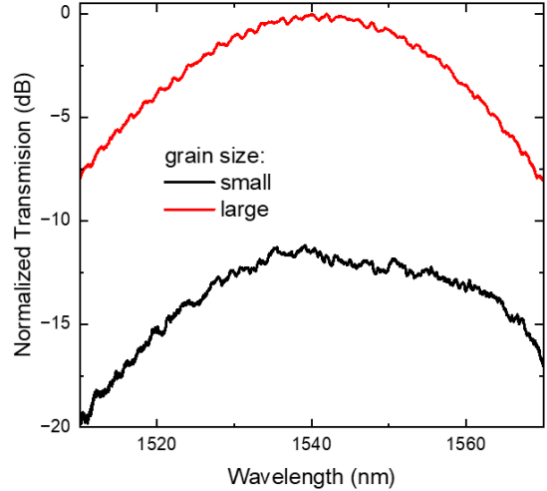

**Fig. S6 The transmission spectra of single mode waveguide with large and small grains.**

**Supplementary Note 4.  Other Phase-change integrated photonic elements characterization**

**Directional couplers**

We wrote directional couplers with various coupling lengths on a phase-change thin film for characterization purposes. The cSb₂Se₃ waveguide used in the couplers has a width of 1.2 μm, and the gap between the coupled waveguides is 350 nm. The measured split ratio between the bar port and the cross port (defined as shown in Fig. S7a) aligns well with the simulation results.

In Fig. S7b, we demonstrate that the split ratio of the directional coupler can be tuned by increasing the gap between the coupled waveguides from 350 nm to 500 nm over a distance of tuning length. Fig. S7d shows the simulation result of the split ratio between the bar and cross ports when changing the tuning length of the directional coupler with a total coupling length of 90 μm. Importantly, this reduction of the waveguide width does not introduce any significant insertion loss (<0.5 dB).

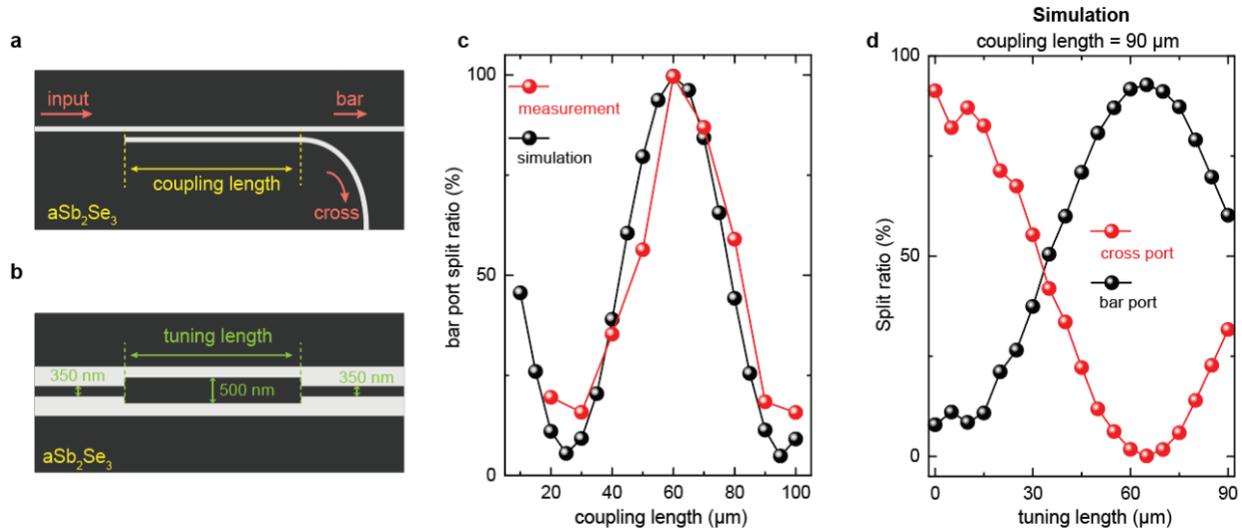

**Figure S7. Sb₂Se₃ directional coupler design. a.** Schematic illustration of the directional coupler calibration measurement. **b.** The zoomed-in schematic of the coupling region showing the coupling of the directional coupler can be tuned by increasing the gap between waveguides from 350 nm to 500 nm. **c.** The bar port transmission ratio as a function of the coupling length. **d**. The split ratio of the directional coupler when tuning the erase length.

## Waveguide Crossing

In conventional dielectric waveguides, such as $Si_3N_4$ waveguides (e.g. 1.2 μm wide, 330 nm thick) on a $SiO_2$ substrate, crosstalk between waveguides becomes non-negligible when they cross over each other. As a result, special waveguide crossing designs are required to minimize the crosstalk and maintain signal integrity. In the case of phase-change waveguides (e.g. 1.2 μm wide, 30 nm thick $cSb_2Se_3$ on top of a $Si_3N_4$ substrate), the crosstalk is significantly reduced even when the two waveguides directly cross over each other over a wide range of cross angles. This is because the phase-change waveguide has a weaker confinement compared to the conventional dielectric waveguide. This makes the optical mode in the $Sb_2Se_3$ waveguide much broader and experiences a weaker index contrast at the waveguide cross. Consequently, no additional design modifications are necessary to mitigate the crosstalk in these waveguide crossings, as shown in our simulation results in Fig. S8.

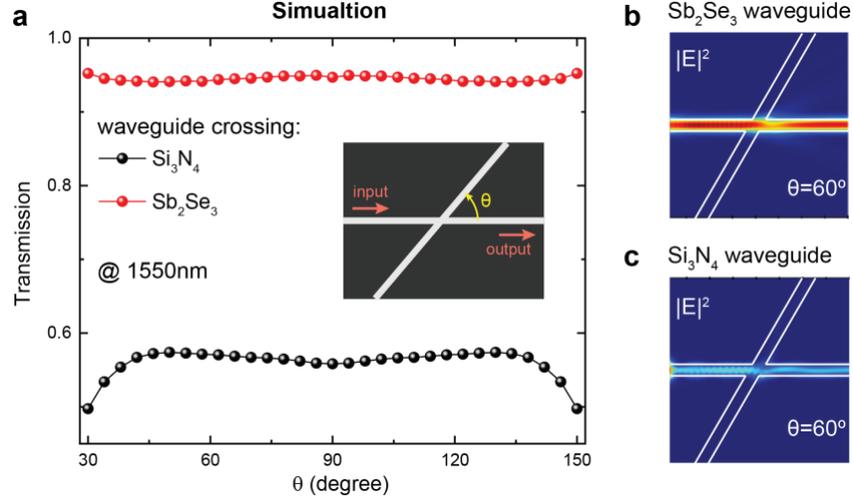

**Figure S8 Sb₂Se₃ waveguide crossing design. a.** Simulation of the transmission of the Sb₂Se₃ and Si₃N4 waveguide crossings. Inset: the schematic of the waveguide crossing geometry. **b and c.** The $|E|^2$ distribution of the Sb₂Se₃ waveguide crossing (**b**) and Si₃N₄ waveguide crossing (**c**) when θ = 60°. The transmission of the Sb₂Se₃ waveguide crossing is much higher than the Si₃N₄ waveguide crossing.

### Inverse-designed waveguide bends

The cSb₂Se₃ waveguide is more sensitive to bending losses compared to conventional dielectric waveguides. This is attributed to the weaker mode confinement and additional scattering caused by randomly oriented crystalline grains. To mitigate bending losses in our photonic devices, we set the bending radius to 120 μm. To further reduce bending losses and enable tighter bending radius, we utilized an inverse-design method to design waveguide bend structures. As depicted in Fig. S9a, these inverse-designed waveguide bends incorporate a DBR-like structure outside the bending region, which helps decrease the loss.

We wrote multiple pairs of single-mode phase-change waveguides, each pair consisting of one waveguide with inverse-designed waveguide bends and another with conventional arc-shape waveguide bends. Both designs featured a bending radius of 35 μm. Our measurements confirm a significant reduction in bending loss by two orders of magnitude when using inverse-designed waveguide bends. It is also worth noting that the conventional arc-shape waveguide bends with a radius of 35 μm exhibited higher bending losses (>10 dB per bend) in the measurement compared to the simulation results (< 0.5 dB per bend). This discrepancy may be attributed to scattering effects caused by the presence of grains within the waveguide, which was not fully captured in the simulations.

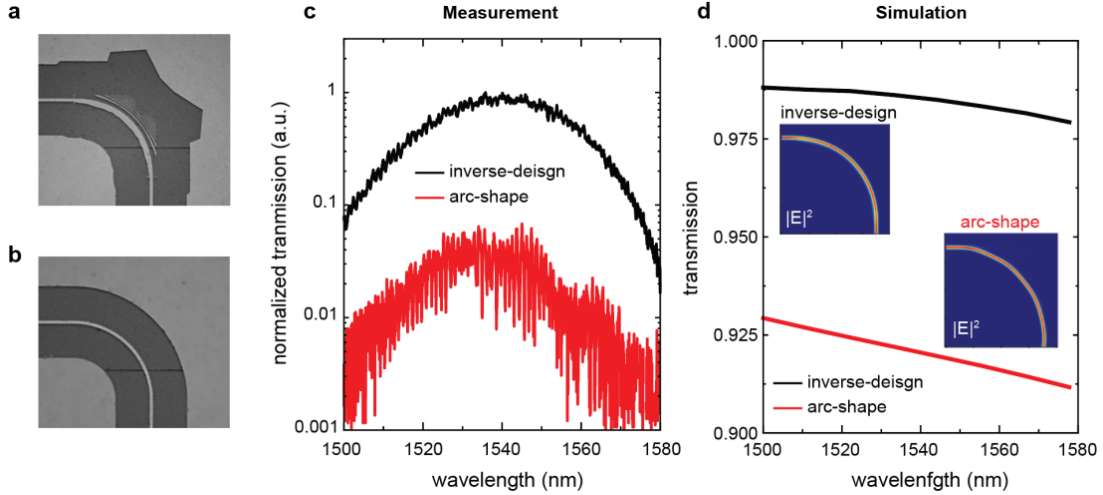

**Figure S9. Inverse-designed and arc-shaped waveguide bends. a** and **b.** Optical images of the inverse-designed (**a**) and arc-shape (**b**) waveguide bends written on phase-change thin film, respectively. Both waveguide bends have a bending radius of 35 μm. **c.** The measured transmission spectra of single mode waveguide consist of one waveguide and two waveguide bends. **d.** The simulated transmission spectra of the waveguide bend. Inset: the mode profile of the inverse-designed waveguide bend and the arc-shape bend at 1550 nm.

## Supplementary Note 5. Tuning the phase-change photonic waveguide

The transmission of the Sb₂Se₃ waveguide can be tuned by selectively erasing or rewriting specific portions of the waveguide. Erasing a section of the waveguide results in a decrease in transmission at the output. Conversely, rewriting a section of the waveguide leads to an increase in transmission. To demonstrate this transmission tuning capability, we erased a 1.2 μm wide waveguide in incremental steps to create a gap up to 30 μm. The corresponding transmission spectra displayed a consistent global decrease in transmission, as illustrated in Fig S10b and S10e. Subsequently, we restored the erased portion of the waveguide, leading to the recovery of the original transmission spectrum and its associated transmission level.

In addition to transmission tuning, the phase response of the waveguide can be selectively tuned by widening specific sections of the waveguide from 1.2 μm to 2 μm. We performed transmission spectra measurements after introducing phase shifts in one arm of the MZI. As increasing the length of the widened waveguide in incremental steps, the interference pattern experienced a gradual redshift, as depicted in Figure S10e. Through fitting analysis of the measurements, we determined that a length of 30 μm of the widened waveguide from 1.2 μm to 2 μm can achieve a π phase shift.

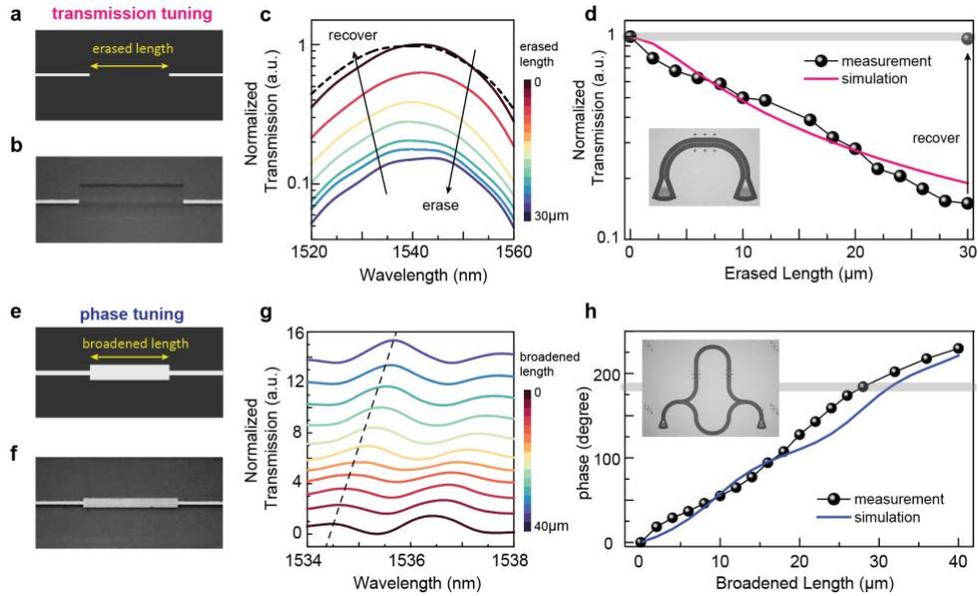

**Figure S10. Tuning the transmission and the phase response of a Sb₂Se₃ waveguide. a.** Schematic of tuning the transmission by varying the erased waveguide length. **b.** Optical image of a partially erased Sb₂Se₃ waveguide. **c.** The transmission spectrum of a Sb₂Se₃ waveguide when an erased length increased from 0 μm to 30 μm in steps. **d.** The waveguide transmission (normalized to the transmission of the pristine Sb₂Se₃ waveguide) under different erased lengths. Inset: the optical image of the waveguide used for transmission measurement. **e.** Schematic of tuning the phase response of a Sb₂Se₃ waveguide by broadening waveguide width. **f.** Optical image of a Sb₂Se₃ waveguide where the center part is broadened. **g.** The shift of transmission spectrum of an MZI when quasi-continuously changing the length of the broadened region in one arm of the MZI from 0 μm to 40 μm. **h.** The induced phase shift under different broadened waveguide lengths. Inset: the optical image of the MZI used for phase-tuning measurement.

**Supplementary Note 6. Demonstration of additional configurations of the reconfigurable photonic interconnect fabric**

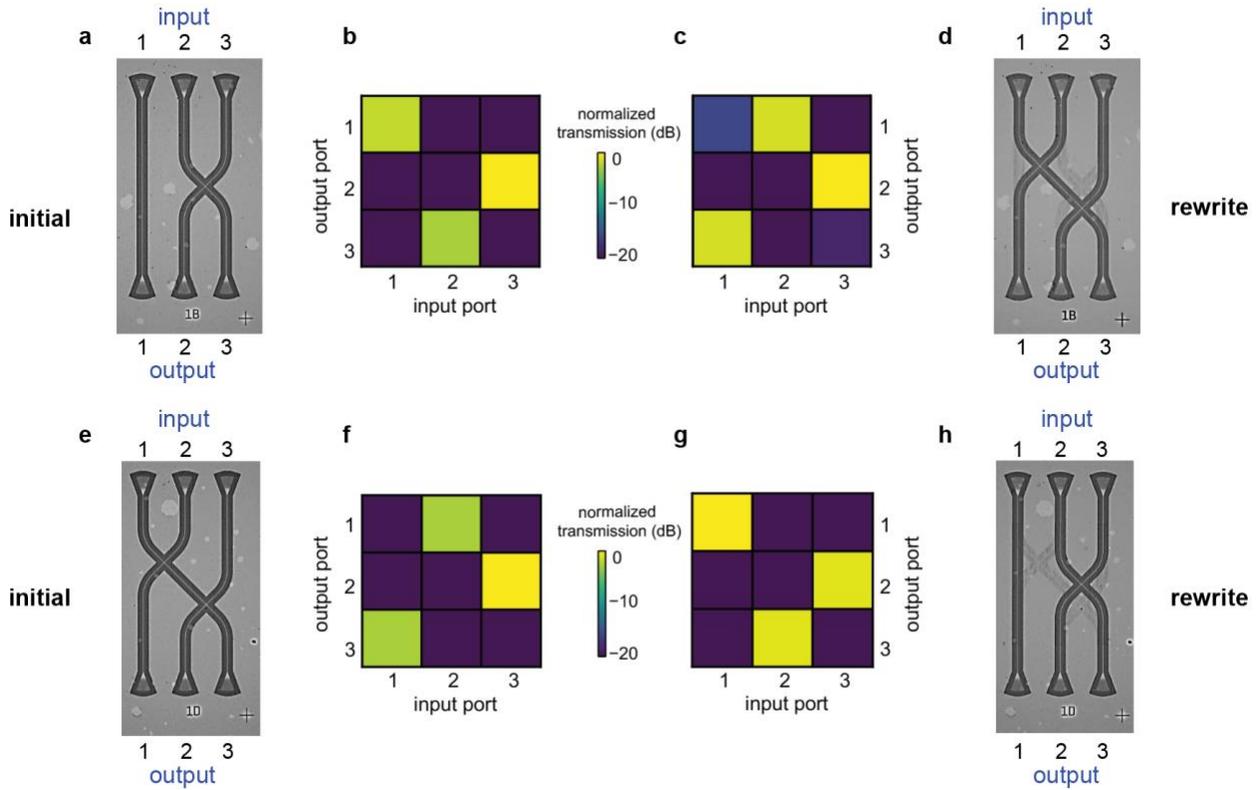

**Figure S11**. **Additional configuration of the reconfigurable photonic interconnect fabric.** The optical image (**a**, **e**), the initial transmission matrices (**b**, **f**), the transmission matrices after reconnection (**c**, **g**), and the optical image after reconnection (**d** and **h**) of two other switches.

## Supplementary Note 7. Modeling the tunable optical filter

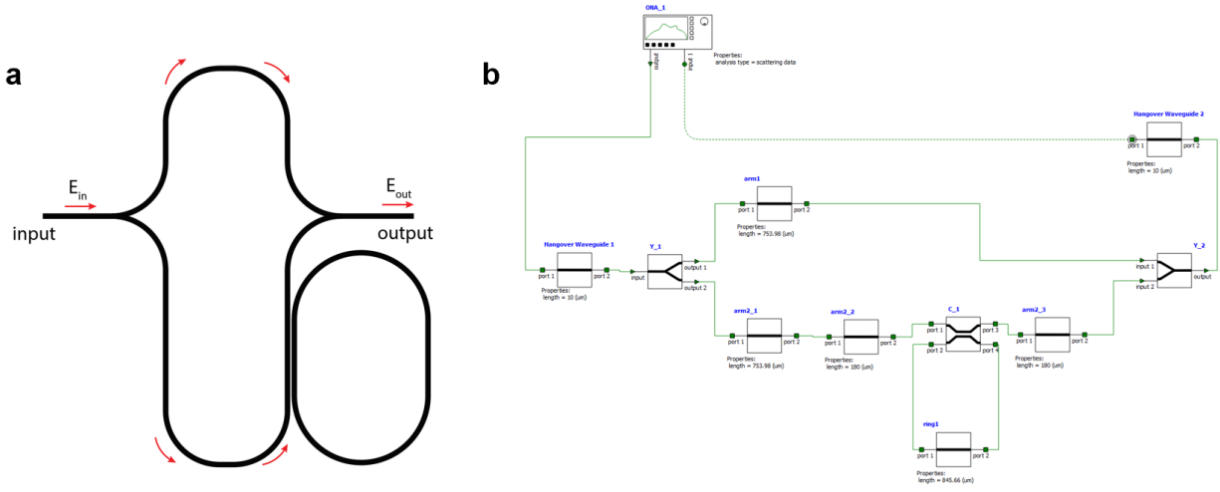

**Figure S12. The ring resonator coupled MZI models. a**. The schematic of the ring resonator coupled MZI. **b**. The ANSYS INTERCONNECT model which is used to simulate the device's performance.

The performance of the optical filter composed of a ring resonator added to an MZI is simulated using ANSYS INTERCONNECT software (Fig. S12). Two devices with different parameters are written and measured experimentally. The response of the devices with the same geometry is also simulated. The dimensions of the devices used in the simulation are the same as the design. Other parameters used in the simulation including the propagation loss, the group index, and so on are obtained from the experimental device characterization. The parameters are listed in Table T2. The simulation results align with the experimental measurement.

**Table T2. Parameters of ring coupled MZIs**

| Device #1 | | Device #2 | |
|---|---|---|---|
| Sb$_2$Se$_3$ waveguide width (μm) | 1.2 | Sb$_2$Se$_3$ waveguide width (μm) | 1.2 |
| Effective index $n_{eff}$ | 1.91 | Effective index $n_{eff}$ | 1.91 |
| Group index $n_{group}$ | 2.47 | Group index $n_{group}$ | 2.47 |
| Propagation loss (dB/mm) | 4.5 | Propagation loss (dB/mm) | 4.5 |
| MZI arm difference $\Delta L$ (μm) | 420 | MZI arm difference $\Delta L$ (μm) | 420 |
| Ring resonators length $L_{Ring}$(μm) | 845.6 | Ring resonators length $L_{Ring}$(μm) | 955.7 |
| Ring & waveguide coupling coefficient | 0.45 | Ring & waveguide coupling coefficient | 0.9 |

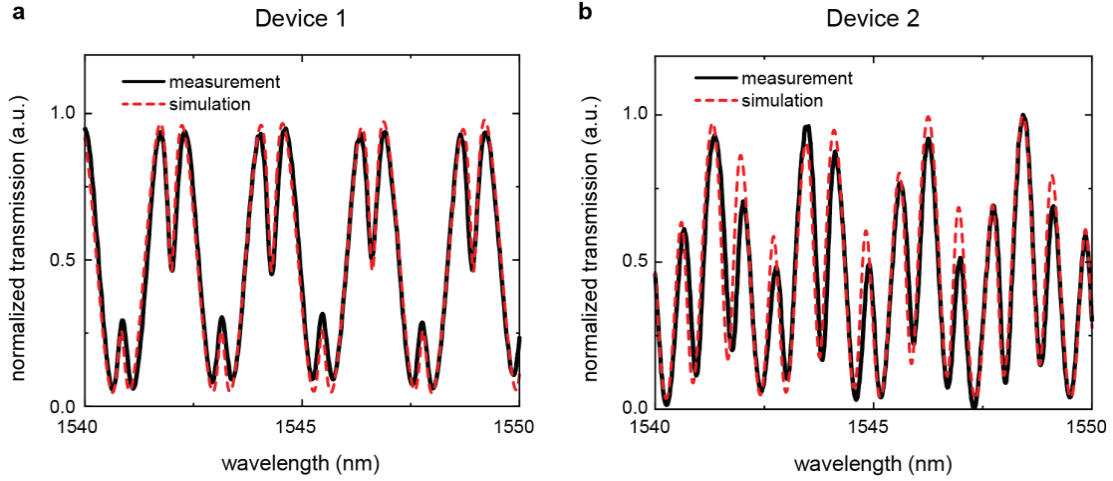

**Figure S13**. **The transmission spectra of ring coupled MZIs. a**. The measured and simulated transmission spectra of device 1. **b**. The measured and simulated transmission spectra of device 2.

Both two filters are then rewritten to perform as the double injection ring filter (DIR). Mathematically, the model describing the transmitted electromagnetic field dependence on the wavelength of the DIR is given by [Cohen, Roei Aviram, Ofer Amrani, and Shlomo Ruschin. *Nature Photonics* **12**.706-712 (2018)]:

$$E_{t1}(\lambda) = \frac{(\tau_1 - \tau_2^* \alpha e^{-i\theta})}{1 - \tau_1^* \tau_2^* \alpha e^{-i\theta}} |E_{i1}(\lambda)| e^{-i\phi_{i1}} - \frac{\kappa_1 \kappa_2^* \sqrt{\alpha} e^{-i\frac{\theta}{2}}}{1 - \tau_1^* \tau_2^* \alpha e^{-i\theta}} |E_{i2}(\lambda)| e^{-i\phi_{i1}},$$

where $\tau = |\tau| e^{-i\varphi_\tau}$, $\kappa = |\kappa| e^{-i\varphi_\kappa}$, $\alpha$, $E_i$, and $\Phi_i$ are the transmission and coupling coefficients of the directional couplers, the loss coefficient of the ring, and the injected fields' intensity and their phase, respectively. $\theta$ is the accumulated phase as the light traverses the ring at a steady state:

$$\theta(\lambda) = \frac{2\pi}{\lambda} n_{eff}(\lambda) L_{Ring},$$

where $n_{eff}$ is the effective index of the propagating mode and $L_{Ring}$ is the perimeter of the ring resonator.

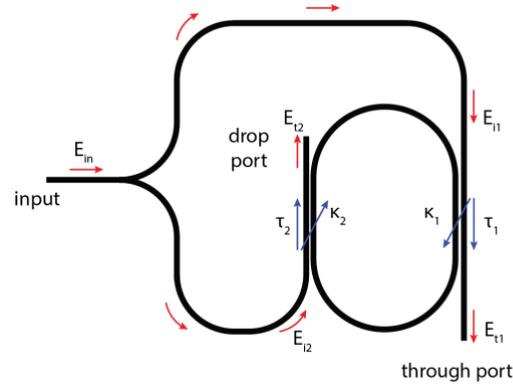

**Figure S14. Schematic illustration of a double infection ring filter.**

The spectral responses of the DIR devices are measured and well-fitted, as shown in Fig. S15a and Fig. S16a. After measurement, we further tune the spectral responses of the DIR devices by modifying the transmission and coupling coefficients of the two DIRs, respectively. Specifically, we tuned the transmission coefficient $\tau_2$ and coupling coefficient $\kappa_2$ in the first DIR (device #1) and tuned the transmission coefficient $\tau_1$ and coupling coefficient $\kappa_1$ in the second DIR (device #2). The spectra responses of the DIRs change accordingly (see Fig. S15b and Fig. S16b). The parameters are listed in Tables T3 and T4.

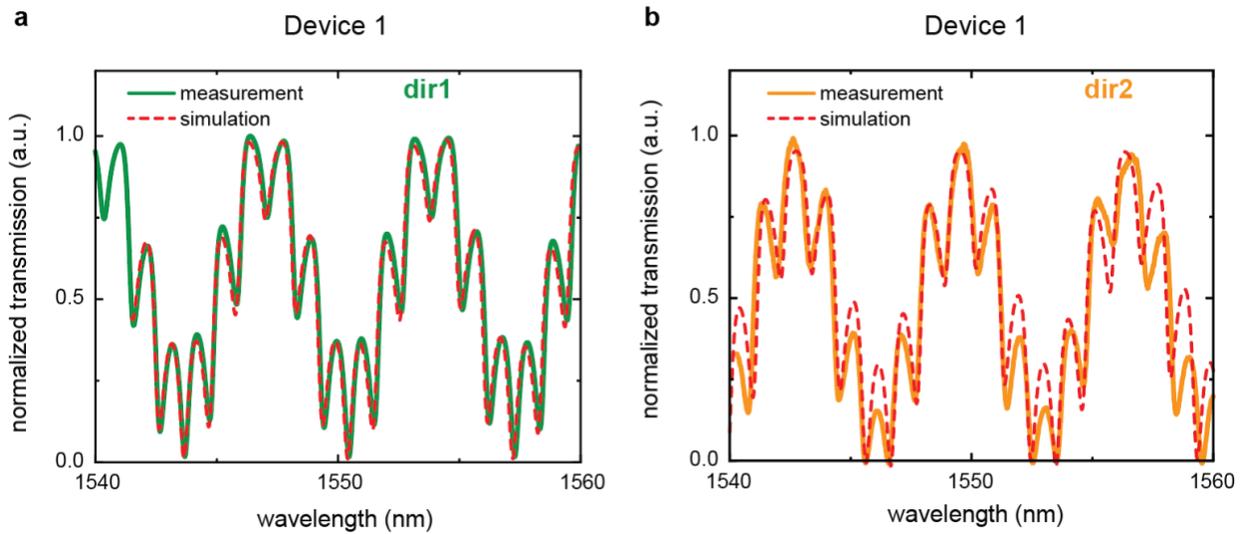

**Figure S15**. **Spectral response of the first double injection ring filters. a.** Measured and simulated spectral response of device #1. **b.** After tuning the transmission coefficient $\tau_2$ and coupling coefficient $\kappa_2$, the spectral response of the same device (dir2) changes.

**Table T3. Fitting Parameters of DIR Device #1**

| Before tuning (dir1) | | After tuning (dir2) | |
|---|---|---|---|
| Coupling coefficient 1 $\lvert\kappa_1\rvert^2$ | 0.45 | Coupling coefficient 1 $\lvert\kappa_1\rvert^2$ | 0.45 |
| Coupling phase 1 $\Phi_{\kappa 1}$ (degree) | 0 | Coupling phase 1 $\Phi_{\kappa 1}$ (degree) | 0 |
| Coupling coefficient 2 $\lvert\kappa_2\rvert^2$ | 0.45 | Coupling coefficient 2 $\lvert\kappa_2\rvert^2$ | 0.4 |
| Coupling phase 1 $\Phi_{\kappa 2}$ (degree) | 0 | Coupling phase 1 $\Phi_{\kappa 2}$ (degree) | 30 |
| Loss coefficient of ring $\alpha$ | 0.45 | Loss coefficient of ring $\alpha$ | 0.45 |
| Injected field intensity 1 $\lvert E_{i1}\rvert^2$ | 50% | Injected field intensity 1 $\lvert E_{i1}\rvert^2$ | 50% |
| Injected field intensity 2 $\lvert E_{i2}\rvert^2$ | 50% | Injected field intensity 2 $\lvert E_{i2}\rvert^2$ | 50% |

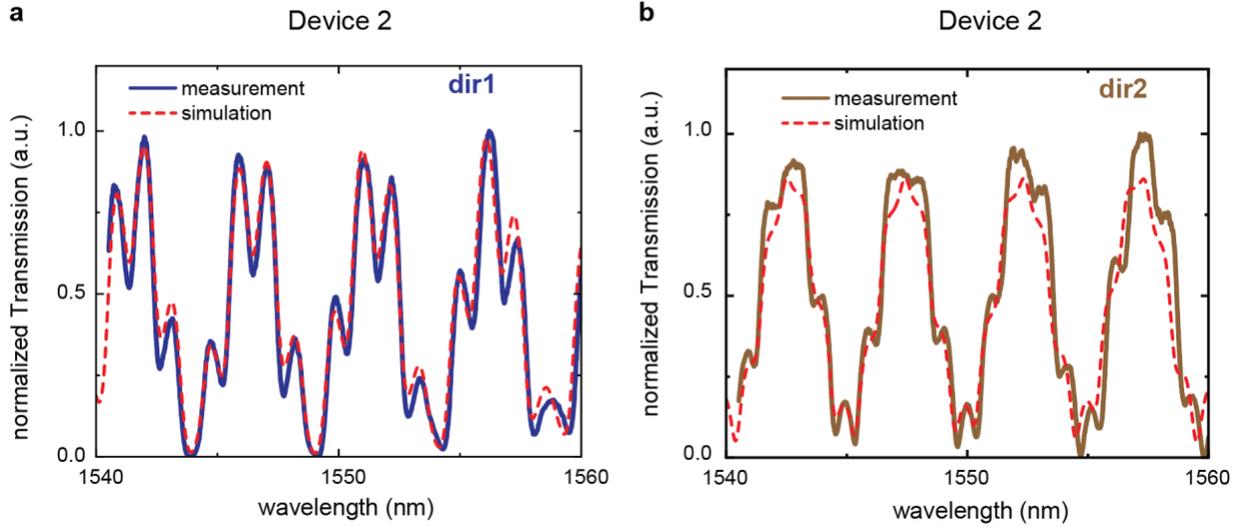

**Figure S16 Spectral response of the second double injection ring filters a.** Measured and simulated spectral response of device #2. **b.** After tuning the transmission coefficient $\tau_1$ and coupling coefficient $\kappa_1$, the spectral response of the same device (dir2) changes.

**Table T4. Fitting Parameters of DIR Device #2**

| Before tuning (dir1) | | After tuning (dir2) | |
|---|---|---|---|
| Coupling coefficient 1 $\lvert\kappa_1\rvert^2$ | 0.9 | Coupling coefficient 1 $\lvert\kappa_1\rvert^2$ | 0.22 |
| Coupling phase 1 $\Phi_{\kappa 1}$ (degree) | 0 | Coupling phase 1 $\Phi_{\kappa 1}$ (degree) | 30 |
| Coupling coefficient 2 $\lvert\kappa_2\rvert^2$ | 0.9 | Coupling coefficient 2 $\lvert\kappa_2\rvert^2$ | 0.9 |
| Coupling phase 1 $\Phi_{\kappa 2}$ (degree) | 0 | Coupling phase 1 $\Phi_{\kappa 2}$ (degree) | 0 |
| Loss coefficient of ring $\alpha$ | 0.45 | Loss coefficient of ring $\alpha$ | 0.45 |
| Injected field intensity 1 $\lvert E_{i1}\rvert^2$ | 50% | Injected field intensity 1 $\lvert E_{i1}\rvert^2$ | 50% |
| Injected field intensity 2 $\lvert E_{i2}\rvert^2$ | 50% | Injected field intensity 2 $\lvert E_{i2}\rvert^2$ | 50% |